\documentclass[aps,prd,twocolumn,floatfix,preprintnumbers,nofootinbib,superscriptaddress]{revtex4}

\usepackage{amsmath}
\usepackage{graphicx}
\usepackage{textcomp}
\usepackage{amsfonts}
\usepackage{amssymb}
\usepackage{bm}
\usepackage[dvips]{color}
\usepackage{multirow}
\usepackage{booktabs}
\usepackage[utf8]{inputenc}

\begin{document}

\title{Investigations on the flavor-dependent axial charges of the octet baryons}

\author{Rui Qi}
\affiliation{School of Physical Science and Technology, Southwest
University, Chongqing 400715, China}

\author{Jin-Bao Wang}
\affiliation{School of Physical Science and Technology, Southwest
University, Chongqing 400715, China}

\author{Gang Li}\email{gli@qfnu.edu.cn}
\affiliation{College of Physics and Engineering, Qufu Normal
University, Qufu 273165, China}

\author{Chun-Sheng An}\email{ancs@swu.edu.cn}
\affiliation{School of Physical Science and Technology, Southwest
University, Chongqing 400715, China}

\author{Cheng-Rong Deng}
\affiliation{School of Physical Science and Technology, Southwest
University, Chongqing 400715, China}

\author{Ju-Jun Xie}
\affiliation{Institute of Modern Physics, Chinese Academy of
Sciences, Lanzhou 730000, China} \affiliation{School of Nuclear
Science and Technology, University of Chinese Academy of Sciences,
Beijing 101408, China} \affiliation{School of Physics and
Microelectronics, Zhengzhou University, Zhengzhou, Henan 450001,
China} \affiliation{Lanzhou Center for Theoretical Physics, Key
Laboratory of Theoretical Physics of Gansu Province, Lanzhou
University, Lanzhou 730000, China}

\thispagestyle{empty}

\date{\today}


\begin{abstract}

We have investigated the axial charges of the ground octet baryons within the extended chiral constituent quark model, where all the possible compact five-quark Fock components $qqq(q\bar{q}) (q=u, d, s)$ in the baryons are considered. The transition couplings between the three- and five-quark
components in the baryons are assumed to be via the $^{3}P_{0}$ mechanism, which could reproduce the sea asymmetry in proton very well. The numerical results for the flavor-dependent axial charges of the octet baryons are comparable to those predicted by other theoretical approaches. It is shown that the singlet axial charges of the octet baryons, which should indicate total baryons spin arising from the spin of the quarks, fall in the range $0.45-0.75$ in present model. This is in consistent with the predictions by lattice QCD and chiral perturbation theory. It's also very interesting that the light quarks spin $\Delta u$ and $\Delta d$ in the $\Lambda$ baryon are of small but negative values, which exactly vanish in the traditional three-quark model.

\end{abstract}

\maketitle


\section{Introduction}
\label{intro}

The flavor-dependent axial charges of baryons, i.e. the singlet axial charge $g^{(0)}_{A}$,
the isovector axial charge $g^{(3)}_{A}$ and the $SU(3)$ octet axial charge $g^{(8)}_{A}$,
are fundamental observables in hadronic physics, since they may provide us information
about the spin structure and properties of baryons. As we know, $g^{(0)}_{A}$ should indicate the total baryons spin
arising from the spin of quarks, and $g^{(3)}_{A}$ and $g^{(8)}_{A}$ should govern the neutron and hyperons $\beta$ decays,
and provide a quantitative measure of spontaneous chiral
symmetry breaking in low energy hadronic physics.
In addition, Goldberger-Treiman relation shows that the isovector axial charge of nucleon
could be directly related to the pion decay constant~\cite{Goldberger:1958tr}.
Therefore, one may establish a connection between the weak and strong interactions by explicit investigations
on the flavor-dependent axial charges.

During the past years from 1990s, intensively experimental measurements on the axial charges of nucleon have
been performed, triggered by the renowned EMC experiments, which raised up the proton spin puzzle~\cite{Ashman:1987hv,Ashman:1989ig}.
Especially, the COMPASS collaboration has made great efforts on corresponding measurements~\cite{COMPASS:2006mhr,COMPASS:2010wkz,COMPASS:2015mhb,COMPASS:2016jwv} at $Q^{2}=3$ ${\rm GeV}^2$, and their latest results show that the singlet axial charge of the proton is $g^{(0)}_{A} = 0.32 \pm 0.02_{\rm stat.} \pm 0.04_{\rm syst.} \pm 0.05_{\rm evol.}$, if the $SU(3)$
flavour symmetry is assumed, which should yield $g^{(8)}_{A}=0.585\pm0.025$~\cite{COMPASS:2016jwv}. And very recently, the first moment $g_{1}$ of nucleon at small $Q^{2}$ is also measured~\cite{COMPASS:2017hef,Deur:2021klh}.
On the other hand, tremendous theoretical investigations on the nucleon spin structure have also been done
using different approaches, such as the lattice QCD ~\cite{Aoki:1996pi,Hagler:2007xi,QCDSF:2011aa,Yang:2016plb,Alexandrou:2017oeh,Yamanaka:2018uud,RQCD:2019jai},
the chiral perturbation theory~\cite{Chen:2001pva,Dorati:2007bk,Lensky:2014dda,Li:2015exr},
the meson cloud models~\cite{Myhrer:2007cf,Thomas:2008ga}, and other phenomenological models with
higher Fock components in nucleon~\cite{Zou:2005xy,An:2005cj,Adamuscin:2007fk,Bijker:2012zza,An:2013daa}.
For recent reviews about the nucleon spin structure, see Refs.~\cite{Kuhn:2008sy,Burkardt:2008jw,Leader:2013jra,Wakamatsu:2014zza,Liu:2015xha,Deur:2018roz,Ji:2020ena}.

For the axial charges of the hyperons, experimental data is still lack up to now.
While the lattice QCD has demonstrated remarkable progress in computing the hyperons
axial transition couplings during the last two decades~\cite{Lin:2007ap,Sasaki:2008ha,Erkol:2009ev,Cooke:2013qqa,Alexandrou:2014vya,Alexandrou:2016xok,Savanur:2018jrb},
and the corresponding calculations are with very high accuracy.
Besides, there have been also other theoretical efforts dedicated to investigations on the hyperons axial
couplings using various approaches, including the
chiral perturbation theory and phenomenological hadronic models
~\cite{Lorce:2007as,Bijker:2009up,Jiang:2009sf,Choi:2010ty,Ledwig:2014rfa,Liu:2018jiu,Jun:2020lfx,Flores-Mendieta:2021wzh}.

Recently, the extended chiral constituent quark model (E$\chi$CQM),
within which higher Fock components in the baryons are included, has been applied to
the sea content of the octet baryons~\cite{An:2012kj}, the sea flavor
asymmetry of the proton $\bar{d}-\bar{u}=0.118\pm0.012$~\cite{NuSea:2001idv,SeaQuest:2021zxb} could be very well
reproduced by taking the model parameters to be the empirical values.
Later, we have applied the work~\cite{An:2012kj} to the baryon sigma terms~\cite{An:2014aea,Duan:2016rkr},
the intrinsic light and strange quark-antiquark pair in the proton and nonperturbative strangeness
suppression~\cite{An:2017flb}, and the orbital angular momentum of the proton~\cite{An:2019tld}.
And in a very recent work, the axial charges of the proton has been investigated
employing the E$\chi$CQM~\cite{Wang:2021ild}. The numerical results obtained
in all the above referred works are in good agreements with predictions by other theoretical
approaches. Consequently, we extend the work~\cite{Wang:2021ild} to estimations on
the hyperons axial charges in present work.

The present paper is organized as follows. In Sec.~\ref{frame},
we give the framework which includes the extended chiral constituent
quark model and the formalism for the octet baryons axial charges in
corresponding model, the explicit numerical results are presented in
Sec.~\ref{num}. Finally, a brief summary is given in
Sec.~\ref{conc}.


\section{Framework}
\label{frame}

In this work, we investigate the axial charges of the ground state octet baryons
employing the E$\chi$CQM, which has been developed to investigate the intrinsic
sea content of the octet baryons in~\cite{An:2012kj}.
Accordingly, in this section, we will briefly introduce the E$\chi$CQM
in Sec.~\ref{ecqm}, and present the formalism for calculations of
octet baryons' axial charges in Sec.~\ref{ac}.

\subsection{E$\chi$CQM}
\label{ecqm}

In the E$\chi$CQM developed in~\cite{An:2012kj}, the wave functions of the ground state
octet baryons can be expressed as:
    \begin{equation}
|B\rangle=\frac{1}{\sqrt{\mathcal{N}}}\left(|qqq\rangle+\sum_{i}C_i^{q}|qqq(q\bar{q}),i\rangle
\right) \label{wfc},
    \end{equation}
where the first term represents the wave function for the
three-quark component of the octet baryons,
and the second term denotes the wave functions for the
compact five-quark components,
with the sum over $i$
runs over all the possible five-quark configurations with a
$q\bar{q}$ ($d\bar{d}$, $u\bar{u}$, $s\bar{s}$, ...)~\footnote{In
this work, we do not take $c\bar{c}$ and $b\bar{b}$ into account,
since the probabilities for them are much smaller than other
five-quark components inside the octet baryons in the low energy scale.}
pair which may form higher Fock components in the octet baryons,
$C_{i}^q/\sqrt{\mathcal{N}}$ are just the corresponding probability
amplitudes for the five-quark components with $\mathcal{N}$
being a normalization constant.

In the present case, we consider the ground states of
baryon octet, whose parities are positive, so that the orbital
quantum number $L$ must be an odd number $1$, $3$, ..., $2n + 1$.
On the other hand, the total spin $S$ of a five-quark system
arising from the quark spin can only be $\frac{1}{2}$, $\frac{3}{2}$,
or $\frac{5}{2}$, therefore, $L$ cannot be higher than $3$ to combine with $S$
to form spin $J_B=L\oplus S$ for the ground state octet baryons.
As discussed explicitly in~\cite{An:2012kj}, only
the five-quark configurations with $L$ = 1 and the $n_r=0$
can form possible Fock components in the ground state octet baryons
with considerable probability amplitudes.

Accordingly, there
are $17$ possible five-quark configurations with different orbital, flavor, spin
and colour symmetries in each of the octet baryons, which are shown explicitly in
Table~\ref{con}, where the symbols $[\dots]_{\nu}$ are the Young tableaux
of corresponding wave functions with $\nu=\chi$, $F$, $S$ and $C$ representing
orbital, flavor, spin and color, respectively. Finally, $[\dots]_{FS}[\dots]_F[\dots]_S$
is just a shorthand for corresponding flavor-spin decomposition.

As we can see in Table~\ref{con}, the $17$ five-quark configurations can be categorized
into two groups according to the spin wave functions of the four-quark subsystem, namely,
spin symmetry $[22]_S$ and $[31]_{S}$, those should lead to the total spin of the four-quark
system $S_4=0$~and~$1$, respectively. For the configurations with spin symmetry $[22]_{S}$,
explicit wave functions can be expressed as
\begin{eqnarray}
|B,\uparrow\rangle_{5q} &=& \sum_{ijkln}\sum_{ab}\sum_{m\bar{s}_z}C^{\frac{1}{2},\uparrow}_{1,m;\frac{1}{2},\bar{s}_z}C^{[1^4]}_{[31]_{\chi FS}^k;[211]_C^{\bar{k}}}C^{[31]_{\chi FS}^{k}}_{[O]_\chi^i;[FS]_{FS}^j} \nonumber \\
&&C^{[FS]_{FS}^{j}}_{[F]_F^l;[22]_{S}^n}C_{a,b}^{[2^{3}]_{C}}|[211]_C^{\bar{k}}(a)\rangle|[11]_{C,\bar{q}}(b)\rangle|I,I_{3}\rangle^{[F]_{F}^{l}} \nonumber \\
&&|1,m\rangle^{[O]_\chi^i}|[22]_S^n\rangle|\bar{\chi},\bar{s}_z\rangle\phi(\{\vec{r}_q\}),
\label{22s}
\end{eqnarray}
where the coefficients $C^{[\cdots]}_{[\cdots][\cdots]}$ represent the CG coefficients of
the $S_{4}$ permutation group, $|[211]^C_{\bar{k}}(a)\rangle$ and $|[11]^{C,\bar{q}}(b)\rangle$
are the color wave functions for the four-quark subsystem and the antiquark, combination of
which should lead to the color singlet $[2^3]_C$.

While combination of the four-quark spin $[31]_{S}$ which leads to $S_4=1$ and the
orbital angular momentum of the five-quark system $L=1$ should result in $J=L\oplus S_4=0$~or~$1$,
both of the two cases of $J$ are legal in present framework, since couplings of both the values of
$J$ and spin of the antiquark $s_{\bar{q}}=1/2$ could lead to $J_{B}=1/2$ for the octet baryons.
Hereafter, we denote these two cases as Set I and
Set II, respectively. And wave functions of the five-quark components octet baryons in these two cases are
\begin{eqnarray}
|B,\uparrow\rangle_{5q} &=& \!\! \sum_{ijkln}\sum_{ab}\sum_{ms_z}C^{00}_{1,m;1,s_z}C^{[1^4]}_{[31]_{\chi FS}^k;[211]_C^{\bar{k}}}C^{[31]_{\chi FS}^{k}}_{[O]^\chi_i;[FS]_{FS}^j}\nonumber \\
&& \!\! C^{[FS]_{FS}^{j}}_{[F]_F^l;[31]_{S}^n}C_{a,b}^{[2^{3}]_{C}}|[211]_C^{\bar{k}}(a)\rangle|[11]_{C,\bar{q}}(b)\rangle|I,I_3\rangle^{[F]_{F}^{l}} \nonumber \\
&& \!\!
|1,m\rangle^{[O]_\chi^i}|[31]_S^n,s_z\rangle|\bar{\chi},\bar{s}_z\rangle\phi(\{\vec{r}_q\}).
\label{31s0}
\end{eqnarray}

\begin{eqnarray}
|B,\uparrow\rangle_{5q} &=& \!\!
\sum_{ijkln}\sum_{ab}\sum_{J_z\bar{s}_z}\sum_{ms_z}
C^{\frac{1}{2},\frac{1}{2}}_{1,J_z;\frac{1}{2},\bar{s}_z}C^{1,J_z}_{1,m;1,s_z}C^{[1^4]}_{[31]_{\chi FS}^k;[211]_C^{\bar{k}}} \nonumber \\
&&\!\! C^{[31]_{\chi FS}^{k}}_{[O]_\chi^i;[FS]_{FS}^j}C^{[FS]_{FS}^{j}}_{[F]_F^l;[31]_{S}^n}C_{a,b}^{[2^{3}]_{C}}|[211]_C^{\bar{k}}(a)\rangle|[11]_{C,\bar{q}}(b)\rangle \nonumber \\
&& \!\! |I,I_3\rangle^{[F]_{F}^{l}}
|1,m\rangle^{[O]^\chi_i}|[31]^S_n,s_z\rangle|\bar{\chi},\bar{s}_z\rangle\phi(\{\vec{r}_q\})\,,
\label{31s1}
\end{eqnarray}
respectively.

One should note that $|I,I_3\rangle^{[F]_{F}^{l}}$ in Eqs. (\ref{22s})-(\ref{31s1})
just represents coupling of the flavour wave functions for the four-quark subsystem
and the antiquark to form appropriate isospin quantum number. Explicit flavor decompositions
of the five-quark configurations in the octet baryons are given in Appendix~\ref{app1}.

{\squeezetable
\begin{table*}[htbp]
\caption{\footnotesize The orbital-flavor-spin configurations for
five-quark configurations those may exist as higher Fock components
in ground octet baryons.} \label{con}
\renewcommand
\tabcolsep{0.10cm}
\renewcommand{\arraystretch}{2}
\scriptsize \vspace{0.7cm}
\begin{tabular}{cccccc}
    \toprule[1.2pt]
    $i$  &  $1$  &  $2$  &  $3$  &  $4$  &  $5$  \\
    Config.&$[31]_{\chi}[4]_{FS}[22]_F[22]_S$&$[31]_{\chi}[31]_{FS}[211]_F[22]_S$&$[31]_{\chi}[31]_{FS}[31]_{F_1}[22]_S$&$[31]_{\chi}[31]_{FS}[31]_{F_2}[22]_S$&$[4]_{\chi}[31]_{FS}[211]_F[22]_S$ \\
\hline

    $i$  &  $6$  &  $7$  &  $8$  &  $9$  &  $10$  \\
    Config.&$[4]_{\chi}[31]_{FS}[31]_{F_1}[22]_S$&$[4]_{\chi}[31]_{FS}[31]_{F_2}[22]_S$&$[31]_{\chi}[4]_{FS}[31]_{F_1}[31]_S$&$[31]_{\chi}[4]_{FS}[31]_{F_2}[31]_S$&$[31]_{\chi}[31]_{FS}[211]_F[31]_S$ \\
\hline

    $i$  &  $11$  &  $12$  &  $13$  &  $14$  &  $15$  \\
    Config.&$[31]_{\chi}[31]_{FS}[22]_F[31]_S$&$[31]_{\chi}[31]_{FS}[31]_{F_1}[31]_S$&$[31]_{\chi}[31]_{FS}[31]_{F_2}[31]_S$&$[4]_{\chi}[31]_{FS}[211]_F[31]_S$&$[4]_{\chi}[31]_{FS}[22]_F[31]_S$ \\
\hline

    $i$  &  $16$  &  $17$  &    &   &    \\
    Config.&$[4]_{\chi}[31]_{FS}[31]_{F_1}[31]_S$&$[4]_{\chi}[31]_{FS}[31]_{F_2}[31]_S$&&& \\
 \bottomrule[1.2pt]
\end{tabular}

\end{table*}
}

The coefficients $C_i^q$ in Eq.~\eqref{wfc}, in the framework of the E$\chi$CQM developed in Ref.~\cite{An:2012kj},
can be related to the coupling between the three-quark and the corresponding five-quark components,
which reads
\begin{equation}
C_i^q=\frac{\langle qqq(q\bar{q}),i|\hat{T}|qqq\rangle}{M_B-E_i}\,,
 \label{coe}
\end{equation}
here $M_B$ denote the physical mass of the baryon $B$~\cite{ParticleDataGroup:2020ssz},
and $E_i$ is the energy
of the $i$-th $qqq(q\bar{q})$ five-quark component.
The transition coupling operator $\hat{T}$ depends on the quark-antiquark creation
mechanism, in present work, we take the widely accepted $^3P_0$ mechanism
following Ref.~\cite{An:2012kj}, where the explicit form of $\hat{T}$ is
\begin{eqnarray}
\hat{T} &=& -\gamma \sum_{j=1,4}
\mathcal{F}_{j,5}^{00}\mathcal{C}_{j,5}^{00}\mathcal{C}_{OFSC}\sum_{m}\langle
1,m;1,-m|00\rangle \times \nonumber \\
&&
\chi_{j,5}^{1,m}\mathcal{Y}_{j,5}^{1,-m}(\vec{p}_j-\vec{p}_5)b^{\dagger}(\vec{p}_j)d^{\dagger}(\vec{p}_5)
\, , \label{3p0}
\end{eqnarray}
where $\gamma$ is an dimensionless transition coupling constant, $\mathcal{F}_{j,5}^{00}\text{ and
}\mathcal{C}_{j,5}^{00}$ are the flavor and color singlet of the
created quark-antiquark pair $q_j\bar{q}_5$, $\chi_{j,5}^{1,m}\text{
and }\mathcal{Y}_{j,5}^{1,-m}$ are the total spin $S_{q\bar{q}}=1$ and relative
orbital $P-$ state of the created quark-antiquark system, the
operator $\mathcal{C}_{OFSC}$ is to calculate the overlap factor
between the residual three-quark configuration in the five-quark
component and the valence three-quark component, finally,
$b^{\dagger}(\vec{p}_j),d^{\dagger}(\vec{p}_5)$ are the quark and
antiquark creation operators.

Finally, to calculate the energy for a given five-quark
configuration $E_{i}$ in Eq.~(\ref{coe}), we employ the
chiral constituent quark model, in which the quark-quark hyperfine
interaction is flavor-dependent~\cite{Glozman:1995fu},
as follow
\begin{eqnarray}
H_{hyp} &=& -\sum_{i<j}\vec{\sigma}_i\cdot
\vec{\sigma}_j\Bigg[\sum_{a=1}^3V_\pi (r_{ij})\lambda^a_i\lambda^a_j
\nonumber\\
&& +\sum_{a=4}^7 V_K (r_{ij})\lambda^a_i\lambda^a_j +V_\eta
(r_{ij})\lambda^8_i\lambda^8_j\Bigg]\, ,
\end{eqnarray}
numerical values for all the exchange coupling
strength constants $V_{M}$ used in present work
are taken to be the empirical
ones~\cite{Glozman:1995fu}. Thus, the energies $E_i$ for the 17
five-quark configurations in Table~\ref{con} should be
\begin{equation}
E_{i}=E_{0}+\langle H_{hyp}\rangle + \delta_{q\bar{q}},
\end{equation}
where $E_{0}$ is a degenerated energy for the 17 five-quark
configurations shown in Table~\ref{con}. The parameter $E_0$ is dependent on the constituent
quark masses, the kinetic quark energies, and also the energies of
the quark confinement interactions and the quark potentials.
Here we take $E_{0}=2127$~MeV,
$\delta_{u\bar{u}} = \delta_{d\bar{d}} = 0$, and $\delta_{s\bar{s}}
= 240$ MeV as used in Ref.~\cite{An:2012kj}.

\subsection{Formalism for the axial charges of octet baryons}
\label{ac}

In this section, we present the formalism for the axial charges
$g_{A}^{(0)}$, $g_{A}^{(3)}$ and $g_{A}^{(8)}$ of octet baryons within the
framework of E$\chi$CQM shown in Sec.~\ref{ecqm}.

The quark spin contribution $\Delta q$, which is related to the flavour-dependent axial vector current operator, is defined as
    \begin{align}
    \langle B,s_z|\int\mathrm{d}x\bar{q}\gamma^{\mu}\gamma^5q|B,s_z\rangle&=s^{\mu}\cdot\Delta q\,,
    \end{align}
where $s^{\mu}$ is the baryon states polarization vector, and $\Delta q$ is
\begin{equation}
\Delta q=(q^{\uparrow}+\bar{q}^{\uparrow})-(q^{\downarrow}+\bar{q}^{\downarrow})\,.
\end{equation}
Combinations of different flavor $\Delta f$ with $f=u,d,s$
lead to flavor-singlet, isovector, and SU(3) octet axial charges of octet baryons, as follows
    \begin{eqnarray}
    g^{(0)}_A &=& \Delta u+\Delta d+\Delta s. \label{g0} \\
    g^{(3)}_A &=& \Delta u-\Delta d, \label{g3} \\
    g^{(8)}_A &=& \Delta u+\Delta d-2\Delta s. \label{g8}
    \end{eqnarray}

In the non-relativistic approximation, $\Delta f$ can be gotten directly through the formula:
    \begin{equation}
    \Delta f=\langle Bs_z|\sum_{j}\,\hat{\sigma}^z_j\delta_{jf}|Bs_z\rangle,
    \end{equation}
where $|Bs_z\rangle$ is the wave function Eq.~(\ref{wfc}) of octet baryons, $\hat{\sigma}^z_j$ is Pauli operator acting on $j$-th quark, and $\delta_{jf}$ is a flavor-dependent operator, defined as
 \begin{eqnarray}
 \delta_{jf}=\begin{cases}1& \text{if the favour of $j$-th quark is $f$}\\0& \text{if the favour of $j$-th quark is not $f$}\end{cases}
 \end{eqnarray}
 Consequently, for each of the octet baryons, one can get
 \begin{eqnarray}
 && \!\!\!\!\! \Delta f = \frac{1}{\mathcal{N}}\langle qqq,s_z|\sum_{j=1,3}\,\hat{\sigma}^z_j\delta_{jf}|qqq,s_z\rangle +\sum_{i}\frac{(C_{i}^{q})^{2}} {\mathcal{N}}  \nonumber \\
 && \!\!\!\!\!\langle
             qqq(q\bar{q}),i,s_z|\sum_{j=1,5}\,\hat{\sigma}^z_j\delta_{jf}|qqq(q\bar{q}),i,s_z\rangle
             ,
             \label{mel}
 \end{eqnarray}
where the non-diagonal terms are neglected.

For simplicity, we denote the matrix elements for the three-quark
components and an explicit given
five-quark component in Eq.~(\ref{mel}) as
    \begin{eqnarray}
    \Delta f^{3q}&=&\langle qqq,s_z|\sum_{j=1,3}\,\hat{\sigma}^z_j\delta_{jf}|qqq,s_z\rangle\,,\\
    \Delta f_i^{5q}&=&\langle qqq(q\bar{q}),i,s_z|\sum_{j=1}^5\,\hat{\sigma}^z_j\delta_{jf}|qqq(q\bar{q}),i,s_z\rangle.\label{5qdelta}
    \end{eqnarray}

Accordingly, one can easily obtain the explicit expression of the matrix
elements results for $\Delta u$, $\Delta d$ and $\Delta s$ of octet baryons
as following
    \begin{equation}
    \Delta u=\frac{1}{\mathcal{N}}\Delta u^{3q}+\sum_{i,q}\frac{(C_{i}^{q})^{2}}{\mathcal{N}}\Delta u_i^{5q}, \label{Deltau}
    \end{equation}
    \begin{equation}
    \Delta d=\frac{1}{\mathcal{N}}\Delta d^{3q}+\sum_{i,q}\frac{(C_{i}^{q})^{2}}{\mathcal{N}}\Delta d_i^{5q}, \label{Deltad}
    \end{equation}
    \begin{equation}
    \Delta s=\frac{1}{\mathcal{N}}\Delta s^{3q}+\sum_{i,q}\frac{(C_{i}^{q})^{2}}{\mathcal{N}}\Delta s_i^{5q}. \label{Deltas}
    \end{equation}

Finally, considering the $SU(2)$ isospin symmetry, one can get
the relations of the spin contributions of quarks with different flavour to spin of the proton
and neutron as
\begin{eqnarray}
&\Delta{u}(p)=\Delta{d}(n),
\Delta{d}(p)=\Delta{u}(n),
\Delta{s}(p)=\Delta{s}(n)\,,
\end{eqnarray}
and the resulted relations of the axial charges of proton and neutron as
\begin{eqnarray}
&g^{(0)}_{A }(p)=g^{(0)}_{A }(n),
g^{(3)}_{A }(p)=-g^{(3)}_{A }(n)\,,\nonumber\\
&g^{(8)}_{A }(p)=g^{(8)}_{A }(n)\,.
\end{eqnarray}
For the spin of quarks with different flavor and the axial charges
of the other octet baryons, analogous relationship can be obtained.


\section{The numerical results and discussions}
\label{num}

In this section, we present our numerical results of the calculations on
the quark spin contributions and the axial charges of the octet baryons in
the framework shown in Sec.~\ref{frame}.

Before going to the final numerical results, firstly, we have to discuss
the model parameters in the E$\chi$CQM. In ref.~\cite{An:2012kj}, the intrinsic
sea contents of the octet baryons have been studied in the E$\chi$CQM by taking
the empirical values for the model parameters, only except for the newly
introduced $V$ which should indicate the transition coupling strength
for the processes $qqq\leftrightarrow qqq(q\bar{q})$, arising from
calculations on the matrix elements of the operator~(\ref{3p0}).
The parameter $V$ is determined by fitting the sea asymmetry in proton
$\bar{d}-\bar{u}=0.118\pm0.012$~\cite{NuSea:2001idv,SeaQuest:2021zxb},
which yields
\begin{equation}
V=570\pm 46~~{\rm MeV}\,.
\end{equation}

As we have discussed in Sec. ~\ref{ecqm}, the spin symmetry of the fourquark
subsystem in the configurations listed in Table~\ref{con} should lead to two sets
for the wave functions of the five-quark components in the octet baryons, namely,
Eqs.~(\ref{31s0})~and~(\ref{31s1}), respectively. In Ref.~\cite{An:2012kj}, only
the former case was considered because of the possible lower energy, while in
the very recent work~\cite{Wang:2021ild}, both of the two sets of wave functions
have been employed to investigate the axial charges of the proton. Keeping all the
other model parameters being the empirical values, the parameter $V$ should be
\begin{equation}
V = 697 \pm {80} ~~{\rm MeV} \, ,
\end{equation}
to fit the intrinsic sea asymmetry of the proton when the set of wave functions~(\ref{31s1})
is used.

In present work, we will consider both Set I and Set II to study the axial charges of
the octet baryons following work~\cite{Wang:2021ild}. And the resulting numerical results for
the probabilities of the five-quark Fock components, the quark spin contributions to
the octet baryons, and the flavor-dependent axial charges of the octet baryons are presented
in the following three subsections, respectively.

\subsection{Probabilities of the intrinsic five-quark Fock components in the octet baryons}
\label{nrp5q}

Employing the values for all the model parameters mentioned above, explicit calculations
on the transition couplings between three- and five-quark components, and energies
for the corresponding five-quark components by taking the wave functions of Set I and Set II
lead to the numerical results for
the probabilities of the intrinsic five-quark components in the octet baryons,
those are shown in Tables~\ref{pdelta}-~\ref{ldelta}.

As discussed in Ref.~\cite{Wang:2021ild}, taking the wave functions of Set II, one can obtain
larger probabilities for the five-quark components in the proton than
those resulted by using the wave functions of Set I. While the results of Set II
are very close to the predictions obtained by the BHPS model~\cite{Chang:2011vx}.
It's very similar for the probabilities of the five-quark in the other octet baryons,
as one can see in the tables.

One may note that the most significant difference of the results obtained in Sets I and II,
is that the probabilities of the five-quark configuration with $i=8$ in all the octet baryons obtained
in Set II are about two times of those obtained in Set I. In fact, one can also
find similar amplifications for the probabilities of the five-quark configurations with $i=9-17$
in those the spin symmetry of the four-quark subsystem is $[31]_{S}$.
These amplifications are mainly caused by the larger matrix elements $\langle \hat{T}\rangle$
for the transition couplings $qqq\leftrightarrow qqq(q\bar{q})$
via the $^{3}P_{0}$ mechanism obtained using the wave function~(\ref{31s1}) in Set II than using~(\ref{31s0})
in Set I.

Straightforwardly, for Set I, one can get the total probabilities of the five-quark components in the octet baryons
\begin{eqnarray}
\mathcal{P}^{N}_{\textrm{I}} = 0.373,~~
\mathcal{P}^{\Sigma}_{\textrm{I}} = 0.338\,,\nonumber\\
\mathcal{P}^{\Xi}_{\textrm{I}} = 0.303,~~
\mathcal{P}^{\Lambda}_{\textrm{I}} = 0.330\, ,
\end{eqnarray}
with uncertainties $\sim10\%$ for all the values caused by the experimental errors of the data for
$\bar{d}-\bar{u}$ in proton, and
\begin{eqnarray}
\mathcal{P}^{N}_{\textrm{II}} = 0.548,~~
\mathcal{P}^{\Sigma}_{\textrm{II}} =0.519 \,,\nonumber  \\
\mathcal{P}^{\Xi}_{\textrm{II}} = 0.467,~~
\mathcal{P}^{\Lambda}_{\textrm{II}} = 0.504\, ,
\end{eqnarray}
with uncertainties $\sim10\%$.

An examination of the obtained probabilities is to estimate the pion- and strangeness-baryon $\sigma$ terms
of the octet baryons. In the model developed in Ref.~\cite{An:2014aea}, the corresponding $\sigma$ terms
could be related to the probabilities of the five-quark components with light and strangeness quark-antiquark
pairs $\mathcal{P}_{q\bar{q}}$ as

\begin{eqnarray}
\label{piN}
\sigma _{\pi N} &=&  \frac{3+2(\mathcal{P}^N_{u \bar u}+\mathcal{P}^N_{d \bar d})}
     {3+2(\mathcal{P}^N_{u \bar u}+\mathcal{P}^N_{d \bar d}-2\mathcal{P}^N_{s \bar s})} \hat{\sigma}, \\
\sigma_{s N}&=& \frac{m_s}{m_l}\frac{2\mathcal{P}_N^{s \bar s}}{3+2(\mathcal{P}_N^{u \bar u}
+\mathcal{P}_N^{d \bar d})}\sigma _{\pi N}\,,
\end{eqnarray}
for nucleon, and
\begin{eqnarray}
 \label{piY}
\sigma_{\pi Y} &=&  \frac{2+ 2 (\mathcal{P}^Y_{u \bar u}+\mathcal{P}^Y_{d \bar d})}
        {3+2(\mathcal{P}^N_{u \bar u}+\mathcal{P}^N_{d \bar d})} \sigma_{\pi N} , \\ [10pt]
\sigma_{s Y} &=&  \frac{1+2\mathcal{P}^Y_{s \bar s}}
          {2 \mathcal{P}^N_{s \bar s}} \sigma_{s N}\nonumber  \\ [10pt]
\label{SY1}
          &=&
          \frac{m_s}{m_l} \frac{1+2\mathcal{P}^Y_{s \bar s}}
          {3+2(\mathcal{P}^N_{u \bar u}+\mathcal{P}^N_{d \bar d})} \sigma_{\pi N}\,,
\label{SY2}
\end{eqnarray}
for the hyperons $\Sigma$ and $\Lambda$, and
\begin{eqnarray}
\sigma_{\pi \Xi} &=& \frac{1+ 2 (\mathcal{P}^\Xi_{u \bar u}+\mathcal{P}^\Xi_{d \bar d})}
        {3+2(\mathcal{P}^N_{u \bar u}+\mathcal{P}^N_{d \bar d})} \sigma_{\pi N}, \\ [10pt]
 \label{piXi}
\sigma_{s \Xi} &=& \frac{2+ 2 \mathcal{P}^\Xi_{s \bar s}}
          {2 \mathcal{P}^N_{s \bar s}} \sigma_{s N}\nonumber  \\ [10pt]
\label{SXi1}
          &=&
          \frac{m_s}{m_l} \frac{2+ 2 \mathcal{P}^\Xi_{s \bar s}}
          {3+2(\mathcal{P}^N_{u \bar u}+\mathcal{P}^N_{d \bar d})} \sigma_{\pi N}\,,
\label{SXi2}
\end{eqnarray}
for the hyperon $\Xi$, where the nonsinglet component
was taken to be $\hat{\sigma}=33\pm5$~MeV, as
extracted within the chiral perturbation theory~\cite{Borasoy:1996bx}
and the ratio of the light and strange quark masses
was taken to be $ \frac{m_s}{m_l}=27.5\pm1$ as shown in PDG~\cite{ParticleDataGroup:2020ssz}.

As shown in~\cite{An:2014aea}, the numerical results of the obtained
employing the wave functions of Set I in present work are in general
consistent with the predictions of chiral perturbation theory and lattice
QCD. While by the obtained probabilities of the five-quark components
shown in Tables~\ref{pdelta}-~\ref{ldelta} employing the wave functions
of Set II, one can get
\begin{eqnarray}
&\sigma_{\pi N}=36\pm5~(\rm MeV),~~~~\sigma_{sN}=46\pm8~(\rm MeV)\,,\\
&\sigma_{\pi\Sigma}=26\pm4~(\rm MeV),~\sigma_{s\Sigma}=302\pm57~(\rm MeV)\,,\\
&\sigma_{\pi\Xi}=16\pm3~(\rm MeV),~\sigma_{s\Xi}=535\pm102~(\rm MeV)\,,\\
&\sigma_{\pi\Lambda}=27\pm4~(\rm MeV),~\sigma_{s\Lambda}=286\pm55~(\rm MeV)\,,
\end{eqnarray}
accordingly, all the results are very close to those obtained in Ref.~\cite{An:2014aea}.

\subsection{The quark spin contributions to the octet baryons}
\label{nrspin}

\begin{table*}[htbp]
\caption{\footnotesize Numerical results for $\Delta{f}$ $(f = u, d, s)$ of the octet baryons, compared to the
predictions by lattice QCD~\cite{QCDSF:2011aa} and chiral effective field theory~\cite{Li:2015exr} shown in the last two
rows, respectively.}
\label{nrdelta}
\renewcommand
\tabcolsep{0.04cm}
\renewcommand{\arraystretch}{2.0}
\scriptsize
\begin{tabular}{c|ccccccccc}
\toprule[1pt]
                              &        &             $p$        &          $n$             &        $\Sigma^+$          &           $\Sigma^0$            &           $\Sigma^-$               &         $\Xi^0$                  &           $\Xi^-$                 &             $\Lambda$    \\

\hline

\multirow{4}{*}{$\Delta{u}$} &  Set I &  $0.883\pm0.005$      &       $-0.213\pm0.003$   &         $0.922\pm0.005$   &           $0.473\pm0.004$      &          $0.023\pm0.003$         &         $-0.215\pm0.002$        &       $0.030\pm0.006$             &   $0.026\pm0.003$     \\

                             &  Set II&  $0.710\pm0.012$      &      $-0.225\pm0.008$    &         $0.762\pm0.014$   &           $0.348\pm0.004$      &          $-0.066\pm0.007$        &         $-0.194\pm0.002$        &       $-0.059\pm0.007$           &    $-0.020\pm0.002$    \\

                             & LQCD~\cite{QCDSF:2011aa}& $0.794(21)(2)$ &  &   &&&  & & \\

                             & $\chi$PT~\cite{Li:2015exr} & $0.90^{+0.03}_{-0.04}$ &  &   &&&  & & \\

\hline

\multirow{4}{*}{$\Delta{d}$} &  Set I &  $-0.213\pm0.003$      &      $0.883\pm0.005$    &         $0.023\pm0.003$   &           $0.473\pm0.004$      &          $0.922\pm0.005$         &         $0.030\pm0.006$        &       $-0.215\pm0.002$             &   $0.026\pm0.003$     \\

                             &  Set II&  $-0.225\pm0.008$      &      $0.710\pm0.012$    &         $-0.066\pm0.007$   &           $0.348\pm0.004$      &          $0.762\pm0.014$        &         $-0.059\pm0.007$        &       $-0.194\pm0.002$           &    $-0.020\pm0.002$    \\

                              & LQCD~\cite{QCDSF:2011aa}& $-0.289(16)(1)$ &  &   &&&  & & \\

                             & $\chi$PT~\cite{Li:2015exr} & $-0.38^{+0.03}_{-0.03}$ &  &  &&&  & & \\

\hline

\multirow{4}{*}{$\Delta{s}$} &  Set I &  $0.015\pm0.002$      &       $0.015\pm0.002$   &         $-0.205\pm0.002$   &           $-0.205\pm0.002$     &          $-0.205\pm0.002$         &         $0.929\pm0.001$        &       $0.929\pm0.001$             &   $0.674\pm0.001$     \\

                             &  Set II&  $-0.020\pm0.003$      &      $-0.020\pm0.003$    &         $-0.180\pm0.003$   &           $-0.180\pm0.003$      &          $-0.180\pm0.003$        &         $0.798\pm0.011$        &       $0.798\pm0.011$           &    $0.551\pm0.006$    \\

                              & LQCD~\cite{QCDSF:2011aa}& $-0.023(10)(1)$ &  &   &&&  & & \\

                             & $\chi$PT~\cite{Li:2015exr} & $-0.007^{+0.004}_{-0.007}$ &  &   &&&  & & \\

\hline
\bottomrule[1pt]

\end{tabular}

\end{table*}

In this section, we present our numerical results for the quark spin contributions to the octet baryons.
Employing the wave functions of Sets I and II, explicit calculations on the matrix elements~(\ref{5qdelta})
lead to numerical results for $\Delta f_{i}^{5q}$ shown in Tables~\ref{pdelta}-~\ref{ldelta} in Appendix~\ref{app2}.

Analogously to the results in Ref.~\cite{Wang:2021ild}, as we can see
in the tables in Appendix~\ref{app2}, the numerical results for $\Delta f_{i}^{5q}$ with $i=1-7$ of Set I are the same with
those of Set II, this is because of that the wave functions of the five-quark configurations $i=1-7$
in Sets I and II are the same one as shown in Eq.~\eqref{22s}. And only spin of the antiquarks contribute to $\Delta f_{i}^{5q}$
of these five-quark configurations, since spin of the fourquark subsystem is $0$. For $\Delta f_{i}^{5q}$ with $i=8-17$,
both the quarks and antiquark should contribute, and the different wave functions of the five-quark configurations in Eqs.~\eqref{31s0}
and~\eqref{31s1} in Sets I and II lead to different results.

Straightforwardly, we can obtain the
numerical results for the quark spin contributions of the octet baryons $\Delta f$ by the
obtained $\Delta f^{5q}_i$ and $\mathcal{P}_{q\bar{q}}$ using Eqs.~(\ref{Deltau})-(\ref{Deltas}),
the results are shown in Table~\ref{nrdelta}.

In most of the theoretical studies on the quark spin contributions of baryons, only nucleon has been considered.
As discussed in Ref.~\cite{Wang:2021ild}, the numerical results of $\Delta f$ of the proton obtained using the
E$\chi$CQM are in general consistent with the predictions by lattice QCD theory. Therefore, here we only compare
the present numerical results to those predicted in Refs.~\cite{QCDSF:2011aa}~and~\cite{Li:2015exr}, a more comprehensive
comparison of the numerical results for $\Delta f$ of the nucleon obtained in present model
to other theoretical predictions can be found in~\cite{Wang:2021ild}.

It's very interesting that $\Delta u$ and $\Delta d$ of the $\Lambda$ hyperon in present model
are small but nonzero values. As we know, in the traditional constituent quark model in which the
baryons are assumed to be composed of three quarks, the two light quarks in $\Lambda$ are completely
antisymmetric, therefore, the $\Lambda$ spin arising from the light quarks spin vanishes. While
in present work, contributions of the five-quark components are taken into account, as we can see
in Table~\ref{ldelta}, several of the five-quark components have nonzero contributions to $\Delta u$ and $\Delta d$.
Accordingly, experimental investigations on the contributions of light quarks to $\Lambda$ spin may be an examination
of the present model.

On the other hand, the magnetic moments, spins and orbital angular momenta of the octet baryons
have been studied in an unquenched quark model in Ref.~\cite{Bijker:2009up}. In that work, the spin of
proton carried by different flavor of quarks were obtained to be $\Delta u=1.098$, $\Delta d=-0.417$
and $\Delta s=-0.005$, and for the $\Lambda$ hyperon, the values $\Delta u=-0.055$, $\Delta d=-0.055$
and $\Delta s=0.961$ were predicted.

\subsection{The flavor-dependent axial charges of the octet baryons}
\label{nrac}

The singlet axial charge $g^{(0)}_A$, isovector axial charge $g^{(3)}_A$ and $SU(3)$ octet axial charge
$g^{(8)}_A$ of the octet baryons can be directly obtained by using Eqs. (\ref{g0})-(\ref{g8}) with the values of $\Delta f$
shown in Table~\ref{nrdelta}, the numerical results are shown in Table~\ref{nrg} in Columns Set I and
Set II for the two sets of wave functions discussed in Sec.~\ref{frame}. As we can see in the
table, the obtained numerical results of $g^{(0)}_A$ of the $\Sigma$ and $\Xi$ hyperons
are larger than that of the nucleon, which should indicate larger contributions of the
quark spin to hyperons spin than those to nucleon spin.

In Ref.~\cite{Wang:2021ild}, the axial charges of the proton obtained using the E$\chi$CQM
have been explicitly compared to the experimental data~\cite{COMPASS:2010wkz}
and predictions of lattice QCD~\cite{Aoki:1996pi,Hagler:2007xi,QCDSF:2011aa,Yang:2016plb,Alexandrou:2017oeh,Yamanaka:2018uud,RQCD:2019jai},
and the chiral perturbation theory~\cite{Chen:2001pva,Dorati:2007bk,Lensky:2014dda,Li:2015exr}.
For the singlet axial charge $g^{(0)}_A$, which should indicate the proton spin arising from
the quarks spin, it was shown that the results in E$\chi$CQM were consistent with predictions of other
theoretical approaches, but larger than the experimental data. In fact, it may be not
convenient for us to directly compare $g^{(0)}_A$ estimated in the static quark model with the
COMPASS data, since $g^{(0)}_A$ is often measured at $Q^{2}=3$~${\rm GeV}^2$,
and the increasing $Q^{2}$ may result in decreased $g^{(0)}_A$~\cite{Jaffe:1987sx}.
Therefore, one may conclude that the estimation of the axial charges of the proton
in E$\chi$CQM are consistent with the data and predictions by lattice QCD and chiral perturbation theory.

Consequently, in present work, we only compare our numerical results for the axial charges of
the other octet baryons with predictions by other theoretical approaches, as we can see in Table~\ref{nrg}.

\begin{table*}[ht]
\caption{\footnotesize Numerical results of $g^{(0)}_{A }$, $g^{(3)}_{A }$, $g^{(8)}_{A }$ of octet baryons, compared to the predictions by other theoretical approaches.}
\label{nrg}
\renewcommand
\tabcolsep{0.1cm}
\renewcommand{\arraystretch}{2.0}
\scriptsize
\begin{tabular}{c|c|cccccccc}
\toprule[1pt]
                              &  Baryon      &             Set I        &         Set II            &        LQCD~\cite{Lin:2007ap}          &           LQCD~\cite{Alexandrou:2016xok}            &           $\chi$PT~\cite{Jiang:2009sf}               &         $\chi$EFT~\cite{Li:2015exr}                  &           RCQM~\cite{Choi:2010ty}                 &             PCQM~\cite{Liu:2018jiu}    \\

\hline

\multirow{4}{*}{$g^{(0)}_{A }$}

                              &  $N$                &  $0.685\pm0.007$      &       $0.465\pm0.023$   &            &           &          &         $0.51^{+0.07}_{-0.08}$        &                   &        \\

                             &  $\Sigma$           &  $0.740\pm0.010$      &      $0.516\pm0.024$    &           &               &  &         &                &        \\

                             & $\Xi$ & $0.744\pm0.009$ & $0.545\pm0.020$ &   &&& & & \\

                             & $\Lambda$ & $0.726\pm0.008$ & $0.511\pm0.010$ &   &&& & & \\

\hline

\multirow{3}{*}{$g^{(3)}_{A }$}
                               &  $N$                &  $1.096\pm0.005$      &       $0.935\pm0.019$   &   $1.18(4)(6)$         &           &          $1.18$   &         $1.27$        &       $1.65$             &   $1.263$     \\

                             &  $\Sigma$           &  $0.899\pm0.008$      &   $0.828\pm0.021$    &         $0.900(42)(54)$   &           $0.762(22)$      &          $1.03$        &                 &       $0.93$    &  $0.896$  \\

                             & $\Xi$ & $-0.245\pm0.008$ & $-0.135\pm0.009$ & $-0.277(15)(19)$  &    $-0.248(9)$&  $-0.23$&  & $-0.32$ & $-0.275$ \\

                             & $\Lambda$ & $0$ & $0$ &   &  $0.085(15)$   &&  $0$& & \\

\hline

\multirow{3}{*}{$g^{(8)}_{A }$}
                              &  $N$                & $0.640\pm0.010$      &       $0.525\pm0.026$  &           &                 &          &         $0.53^{+0.06}_{-0.06}$        &                    &        \\

                             &  $\Sigma$           &  $1.355\pm0.011$      &  $1.056\pm0.026$     &            &               &          &         &                  &        \\

                             & $\Xi$ & $-2.043\pm0.010$ & $-1.849\pm0.031$ &   &&& & & \\

                             & $\Lambda$ & $-1.296\pm0.009$ & $-1.142\pm0.016$ &   &&&  & & \\

\hline
\bottomrule[1pt]

\end{tabular}

\end{table*}

In Ref.~\cite{Lin:2007ap}, the first lattice calculations on the axial charge $g^{(3)}_A$
of the $\Sigma$ and $\Xi$ hyperons was performed using $2+1$-flavor lattices by Lin and Orginos.
And in 2018, Savanur and Lin presented the first chiral-continuum–finite-volume extrapolation of
the hyperons axial couplings $g_{\Sigma\Sigma}$ and $g_{\Xi\Xi}$ from $N_f=2+1+1$ lattice QCD
in~\cite{Savanur:2018jrb}, where the axial charges $g^{(3)}_A$ of $\Sigma$ and $\Xi$ with respect
to the nucleon axial charges were predicted to be $g^{(3)}_A/g_A=0.702(12)(4)$ for $\Sigma$ and
$g^{(3)}_A/g_A=-0.213(5)(1)$ for $\Xi$, respectively.

In Ref.~\cite{Erkol:2009ev}, the strangeness-conserving $NN$, $\Sigma\Sigma$, $\Xi\Xi$, $\Lambda\Sigma$ and
the strangeness-changing $\Lambda N$, $\Sigma N$, $\Lambda\Xi$, $\Sigma\Xi$ axial charges
was studied in lattice QCD with $N_f=2$. The resulting axial charges of the $\Sigma$ and
$\Xi$ fall in the same range as the present obtained numerical results.

And in a recent work~\cite{Alexandrou:2016xok}, the axial couplings of the low lying baryons were
evaluated using a total of five ensembles of dynamical
twisted mass fermion gauge configurations. As we can see in Table~\ref{nrg},
the present numerical results for the axial charges $g^{(3)}_A$ and $g^{(8)}_A$ of the
$\Sigma$, $\Xi$ and $\Lambda$ hyperons are very close to those depicted in the
figures in Ref.~\cite{Alexandrou:2016xok}, and in Table~\ref{nrg}, we only show their predicted
values for $g^{(3)}_A$ of the $\Sigma$ and $\Xi$ hyperons.

The axial charges of the octet baryons have also been investigated using chiral perturbation
theory~\cite{Li:2015exr,Jiang:2009sf}. In~\cite{Li:2015exr}, all the three flavor-dependent
axial charges of the nucleon has been studied, as shown in the column $\chi$EFT
in Table~\ref{nrg}.
In Ref.~\cite{Jiang:2009sf}, $g^{(3)}_A$ of the nucleon was predicted to be $1.18$, while that for the
hyperons $\Sigma$
and $\Xi$ were predicted to be $1.03$ and $-0.23$, respectively, as we can see in the column $\chi$PT
in Table~\ref{nrg}.

In addition, we also compare the present obtained results with those obtained using
the relativistic constituent $qqq$ quark model~\cite{Choi:2010ty} and perturbative
chiral quark model~\cite{Liu:2018jiu}, which are shown in the columns RCQM and PCQM in
Table~\ref{nrg}, respectively.


\section{Summary}
\label{conc}

To summarize, we investigate the intrinsic five-quark components and the flavor-dependent
axial charges of the octet baryons within the framework of the extended chiral constituent
quark model, in which wave functions of the higher five-quark Fock components in baryons
are taken into account.

The probabilities of the five-quark components in the nucleon are in agreements with other theoretical predictions.
And with the obtained probabilities of the five-quark components
in the wave functions of the octet baryons, we get the pion- and strangeness-baryon sigma
terms consistent with predictions by lattice QCD.

Finally, the present obtained numerical results show that the singlet axial charges of the octet baryons,
which should indicate total baryons spin arising from the spin of the quarks, fall in the
range $0.45-0.75$ in present model, it's in consistent with the predictions by lattice QCD,
chiral perturbation theory, as well the other theoretical approaches.
The present obtained axial charges $g^{(3)}_A$ and $g^{(8)}_A$ of the octet baryons are
also comparable to those predicted by lattice QCD and chiral perturbation theory.
It's also very interesting that the quark spin $\Delta u$ and $\Delta d$ of $\Lambda$
arising from the five-quark components are of small
but nonzero values.


\begin{acknowledgments}

This work is partly supported by the Chongqing Natural Science
Foundation under Project No. cstc2021jcyj-msxmX0078, and
No. cstc2019jcyj-msxmX0409,
and the National Natural Science Foundation of China under Grant Nos.
12075288, 12075133, 11735003, 11961141012 and 11835015. It is also
supported by
the Youth Innovation Promotion Association CAS, Taishan
Scholar Project of Shandong Province (Grant No.tsqn202103062),
the Higher Educational Youth Innovation Science and Technology
Program Shandong Province (Grant No. 2020KJJ004).

\end{acknowledgments}



\begin{appendix}

\setcounter{table}{0}
\renewcommand\thetable{A\Roman{table}}

\section{Flavor decompositions of five-quark components in the octet baryons}
\label{app1}
In each five-quark configuration, $[v]_F$ is the flavour wave function of four-quark subsystem, one can get the flavour wave function of the five-quark system by combining $[v]_F$ and antiquark flavour wave function $|\bar{q}\rangle$.
Here we give the explicit flavor decompositions for all the possible five-quark configurations
in the octet baryons.

\subsection{Nucleon}

For the proton, whose isospin wave function
is $|\frac{1}{2},\frac{1}{2}\rangle_{I}$,
the configurations with $[v]_F=[31]_{F_1}$ rules out the strangeness component.
Then the four-quark subsystem should couple with the antiquark as
\begin{eqnarray}
|\frac{1}{2},\frac{1}{2}\rangle_{I}^{[31]_{F_1}} &=&
\sqrt{\frac{2}{3}}|u^3d_{[31]_{F_1}}\rangle\otimes|\bar{u}\rangle +
\nonumber \\
&&
\sqrt{\frac{1}{3}}|u^2d^2_{[31]_{F_1}}\rangle\otimes|\bar{d}\rangle,
\label{31f1}
\end{eqnarray}
For the configurations with $[v]_F=[31]_{F_2}$, the
quark-antiquark pair can only be $s\bar{s}$, and the corresponding
isospin wave function of the proton is
\begin{equation}
|\frac{1}{2},\frac{1}{2}\rangle_{I}^{[31]_{F_2}}=|u^2ds_{[31]_{F_2}}\rangle\otimes|\bar{s}\rangle,
\label{31f2}
\end{equation}
For the configurations with $[v]_F=[22]_F$, the quark-antiquark pair
can be $d\bar{d}$ and $s\bar{s}$. To take into account the $SU(3)$
flavour symmetry breaking effects, we treat these two kinds of
$qqq(q\bar{q})$ configurations separately. And the flavour decompositions
are
\begin{eqnarray}
|\frac{1}{2},\frac{1}{2}\rangle_{I}^{[22]_{F}}&=&|u^2d^2_{[22]_F}\rangle\otimes|\bar{d}\rangle, \\
|\frac{1}{2},\frac{1}{2}\rangle_{I}^{[22]_{F}}&=&|u^2ds_{[22]_F}\rangle\otimes|\bar{s}\rangle\,,
\end{eqnarray}

At last, for the configurations with $[v]_F=[211]_F$,
which limits the quark-antiquark pair to be $s\bar{s}$, the flavour decomposition is
\begin{equation}
|\frac{1}{2},\frac{1}{2}\rangle_{I}^{[211]_{F}}=|u^2ds_{[211]_{F}}\rangle\otimes|\bar{s}\rangle
. \label{211f}
\end{equation}

Analogously, one can get the following flavour decompositions for the
five-quark configurations in neutron according to the isospin quark
number:
\begin{eqnarray}
|\frac{1}{2},-\frac{1}{2}\rangle_{I}^{[31]_{F_1}} &=&
\sqrt{\frac{1}{3}}|u^2d^2_{[31]_{F_1}}\rangle\otimes|\bar{u}\rangle +
\nonumber \\
&&
\sqrt{\frac{2}{3}}|ud^3_{[31]_{F_1}}\rangle\otimes|\bar{d}\rangle\,,\\
|\frac{1}{2},-\frac{1}{2}\rangle_{I}^{[31]_{F_2}}&=&|ud^2s_{[31]_{F_2}}\rangle\otimes|\bar{s}\rangle\,,\\
|\frac{1}{2},-\frac{1}{2}\rangle_{I}^{[22]_{F}}&=&-|u^2d^2_{[22]^F}\rangle\otimes|\bar{u}\rangle\,,\\
|\frac{1}{2},-\frac{1}{2}\rangle_{I}^{[22]_{F}}&=&|ud^2s_{[22]^F}\rangle\otimes|\bar{s}\rangle\,,\\
|\frac{1}{2},-\frac{1}{2}\rangle_{I}^{[211]_{F}}&=&|ud^2s_{[211]_{F}}\rangle\otimes|\bar{s}\rangle\,.
\end{eqnarray}

\subsection{$\Sigma$ baryons}

For the $\Sigma^+$ baryon, the isospin wave function
is $|1,1\rangle_{I}$.
In the five-quark configurations with $[v]_{F}=[31]_{F_1}$,
both the light and strange quark-antiquark pairs survive. The corresponding
flavour decompositions are
\begin{eqnarray}
|1,1\rangle_{I}^{[31]_{F_1}} &=&
\sqrt{\frac{3}{4}}|u^3s_{[31]_{F_1}}\rangle\otimes|\bar{u}\rangle +
\nonumber \\
&&
\sqrt{\frac{1}{4}}|u^2ds_{[31]_{F_1}}\rangle\otimes|\bar{d}\rangle\,,\\
|1,1\rangle_{I}^{[31]_{F_1}}&=&|u^2s^2_{[31]_{F_1}}\rangle\otimes|\bar{s}\rangle\,,
\end{eqnarray}
respectively.

For the configurations with $[v]_F=[31]_{F_2}$, in present case,
flavour decomposition of the $uuss\bar{s}$ configuration is the same as
the configurations with $[v]_F=[31]_{F_1}$.
And the flavour decomposition of the five-quark configurations with
light quark antiquark pair should be
\begin{equation}
|1,1\rangle_{I}^{[31]_{F_2}}=|u^2ds_{[31]_{F_2}}\rangle\otimes|\bar{d}\rangle\,.
\end{equation}

For the configurations with $[v]_F=[22]_F$, both light and strangeness five-quark components exist,
and the flavor decompositions are
\begin{eqnarray}
|1,1\rangle_{I}^{[22]_{F}}&=&-|u^2ds_{[22]^F}\rangle\otimes|\bar{d}\rangle, \\
|1,1\rangle_{I}^{[22]_{F}}&=&|u^2s^2_{[22]^F}\rangle\otimes|\bar{s}\rangle,
\end{eqnarray}

Finally, the configurations with $[v]_F=[211]_F$ rules out the strangeness
five-quark component in $\Sigma^{+}$, only $uuds\bar{d}$ component exists. And
the flavour decomposition reads
\begin{equation}
|1,1\rangle_{I}^{[211]_{F}}=-|u^2ds_{[211]_{F}}\rangle\otimes|\bar{d}\rangle\,.
\end{equation}

Considering the isospin $SU(2)$ symmetry, one can obtain the flavour decomposition
\begin{eqnarray}
|1,0\rangle_{I}^{[31]_{F_1}} &=&
\sqrt{\frac{1}{2}}|u^2ds_{[31]_{F_1}}\rangle\otimes|\bar{u}\rangle -
\nonumber \\
&&
\sqrt{\frac{1}{2}}|ud^2s_{[31]_{F_1}}\rangle\otimes|\bar{d}\rangle\,,\\
|1,0\rangle_{I}^{[31]_{F_1}}&=&|uds^2_{[31]_{F_1}}\rangle\otimes|\bar{s}\rangle\,,\\
|1,0\rangle_{I}^{[31]_{F_2}} &=&
\sqrt{\frac{1}{2}}|u^2ds_{[31]_{F_2}}\rangle\otimes|\bar{u}\rangle -
\nonumber \\
&&
\sqrt{\frac{1}{2}}|ud^2s_{[31]_{F_2}}\rangle\otimes|\bar{d}\rangle\,,\\
|1,0\rangle_{I}^{[22]_F} &=&
-\sqrt{\frac{1}{2}}|u^2ds_{[22]_F}\rangle\otimes|\bar{u}\rangle +
\nonumber \\
&&
\sqrt{\frac{1}{2}}|ud^2s_{[22]_F}\rangle\otimes|\bar{d}\rangle\,,\\
|1,0\rangle_{I}^{[22]_{F}}&=&|uds^2_{[22]_F}\rangle\otimes|\bar{s}\rangle\,,\\
|1,0\rangle_{I}^{[211]_{F_2}} &=&
\sqrt{\frac{1}{2}}|u^2ds_{[211]_F}\rangle\otimes|\bar{u}\rangle -
\nonumber \\
&&
\sqrt{\frac{1}{2}}|ud^2s_{[211]_F}\rangle\otimes|\bar{d}\rangle\,,
\end{eqnarray}
for the five-quark components in the $\Sigma^{0}$ baryon, and
\begin{eqnarray}
|1,-1\rangle_{I}^{[31]_{F_1}} &=&
\sqrt{\frac{1}{4}}|ud^2s_{[31]_{F_1}}\rangle\otimes|\bar{u}\rangle +
\nonumber \\
&&
\sqrt{\frac{3}{4}}|d^3s_{[31]_{F_1}}\rangle\otimes|\bar{d}\rangle\,,\\
|1,-1\rangle_{I}^{[31]_{F_1}}&=&|d^2s^2_{[31]_{F_1}}\rangle\otimes|\bar{s}\rangle\,,\\
|1,-1\rangle_{I}^{[31]_{F_2}}&=&|ud^2s_{[31]_{F_2}}\rangle\otimes|\bar{u}\rangle\,,\\
|1,-1\rangle_{I}^{[22]_{F}}&=&-|ud^2s_{[22]^F}\rangle\otimes|\bar{u}\rangle\,,\\
|1,-1\rangle_{I}^{[22]_{F}}&=&|d^2s^2_{[22]^F}\rangle\otimes|\bar{s}\rangle\,,\\
|1,-1\rangle_{I}^{[211]_{F}}&=&|ud^2s_{[211]_{F}}\rangle\otimes|\bar{u}\rangle\,,
\end{eqnarray}
for the five-quark components in the $\Sigma^{-}$ baryon, respectively.

\subsection{$\Xi$ baryon}

The isospin wave function of the $\Xi^+$ baryon is $|1/2,1/2\rangle_{I}$.
Accordingly, for the configurations with $[v]_{F}=[31]_{F_1}$,
both the five-quark components with light and strange quark-antiquark
pairs can exist. Accordingly, the flavour decompositions are
\begin{eqnarray}
|\frac{1}{2},\frac{1}{2}\rangle_{I}^{[31]_{F_1}} &=&
\sqrt{\frac{2}{3}}|u^2s^2_{[31]_{F_1}}\rangle\otimes|\bar{u}\rangle +
\nonumber \\
&&
\sqrt{\frac{1}{3}}|uds^2_{[31]_{F_1}}\rangle\otimes|\bar{d}\rangle\,,\\
|\frac{1}{2},\frac{1}{2}\rangle_{I}^{[31]_{F_1}}&=&|us^3_{[31]_{F_1}}\rangle\otimes|\bar{s}\rangle,
\end{eqnarray}
respectively.

For the configurations with $[v]_F=[31]_{F_2}$, the $usss\bar{s}$ component is the same as the one with $[31]_{F_1}$. So we only have to consider the light
five-quark components, which should be
\begin{equation}
|\frac{1}{2},\frac{1}{2}\rangle_{I}^{[31]_{F_2}}=|uds^2_{[31]_{F_2}}\rangle\otimes|\bar{d}\rangle\,.
\end{equation}

The flavour symmetry $[v]_F=[22]_F$ rules out the five-quark component in
$\Xi^{+}$ with strange quark-antiquark pair, and
the flavour decomposition for the $uss(q\bar{q})$ component is
\begin{eqnarray}
|\frac{1}{2},\frac{1}{2}\rangle_{I}^{[22]_{F}} &=&
-\sqrt{\frac{2}{3}}|u^2s^2_{[22]_{F}}\rangle\otimes|\bar{u}\rangle -
\nonumber \\
&&
\sqrt{\frac{1}{3}}|uds^2_{[22]_{F}}\rangle\otimes|\bar{d}\rangle\,.
\end{eqnarray}

Finally, for the configurations with $[v]_F=[211]_F$, only $udss\bar{d}$ component exists in
$\Xi^{+}$. And
\begin{equation}
|\frac{1}{2},\frac{1}{2}\rangle_{I}^{[211]_{F}}=-|uds^2_{[211]_{F}}\rangle\otimes|\bar{d}\rangle\,.
\end{equation}

The flavour decompositions of the five-quark components in the $\Xi^{-}$
baryon can be obtained by considering the $SU(2)$ isospin symmetry as
\begin{eqnarray}
|\frac{1}{2},-\frac{1}{2}\rangle_{I}^{[31]_{F_1}} &=&
\sqrt{\frac{1}{3}}|uds^2_{[31]_{F_1}}\rangle\otimes|\bar{u}\rangle +
\nonumber \\
&&
\sqrt{\frac{2}{3}}|d^2s^2_{[31]_{F_1}}\rangle\otimes|\bar{d}\rangle\,,\\
|\frac{1}{2},-\frac{1}{2}\rangle_{I}^{[31]_{F_1}}&=&|ds^3_{[31]_{F_1}}\rangle\otimes|\bar{s}\rangle\,,\\
|\frac{1}{2},-\frac{1}{2}\rangle_{I}^{[31]_{F_2}}&=&|uds^2_{[31]_{F_2}}\rangle\otimes|\bar{u}\rangle\,,\\
|\frac{1}{2},-\frac{1}{2}\rangle_{I}^{[22]_{F}} &=&
-\sqrt{\frac{1}{3}}|uds^2_{[22]_{F}}\rangle\otimes|\bar{u}\rangle -
\nonumber \\
&&
\sqrt{\frac{2}{3}}|d^2s^2_{[22]_{F}}\rangle\otimes|\bar{d}\rangle\,,\\
|\frac{1}{2},-\frac{1}{2}\rangle_{I}^{[211]_{F}}&=&|uds^2_{[211]_{F}}\rangle\otimes|\bar{u}\rangle\,.
\end{eqnarray}

\subsection{$\Lambda$ baryon}

In the $\Lambda$ baryon, isospin zero rules out the five-quark configurations
with $[v]_{F}=[31]_{F_1}$.
For the configurations with $[v]_{F}=[31]_{F_2}$, the light five-quark components should be
\begin{eqnarray}
|0,0\rangle_{I}^{[31]_{F_2}} &=&
\sqrt{\frac{1}{2}}|u^2ds_{[31]_{F_2}}\rangle\otimes|\bar{u}\rangle +
\nonumber \\
&&
\sqrt{\frac{1}{2}}|ud^2s_{[31]_{F_2}}\rangle\otimes|\bar{d}\rangle\,.
\end{eqnarray}
Additionally, the strangeness five-quark component can also exist in $\Lambda$, as
\begin{equation}
|0,0\rangle_{I}^{[31]_{F_2}}=|uds^2_{[31]_{F_2}}\rangle\otimes|\bar{s}\rangle\,.
\end{equation}

For the five-quark configurations with $[v]_F=[22]_F$, only the ones with light
quark-antiquark pairs exist in $\Lambda$ baryon, and the flavour decomposition is
\begin{eqnarray}
|0,0\rangle_{I}^{[22]_F} &=&
-\sqrt{\frac{1}{2}}|u^2ds_{[22]_F}\rangle\otimes|\bar{u}\rangle -
\nonumber \\
&&
\sqrt{\frac{1}{2}}|ud^2s_{[22]_F}\rangle\otimes|\bar{d}\rangle.
\end{eqnarray}

Finally, for the configurations with $[v]_F=[211]_F$, the light five-quark components should be
\begin{eqnarray}
|0,0\rangle_{I}^{[211]_{F_2}} &=&
-\sqrt{\frac{1}{2}}|u^2ds_{[211]_F}\rangle\otimes|\bar{u}\rangle -
\nonumber \\
&&
\sqrt{\frac{1}{2}}|ud^2s_{[211]_F}\rangle\otimes|\bar{d}\rangle\,,
\end{eqnarray}
and the strangeness component is
\begin{equation}
|0,0\rangle_{I}^{[211]_{F}}=|uds^2_{[211]_{F}}\rangle\otimes|\bar{s}\rangle\,.
\end{equation}

\section{Numerical results of each five-quark components $\mathcal{P}{q\bar{q}}$ and $\Delta{f^q_i}$$(f=u, d, s)$ of octet baryons.}
\label{app2}

In present appendix, we show the numerical results on the matrix elements of
$\Delta{f^q_i}$$(f=u, d, s)$ for the octet baryons, as shown in Tables~\ref{pdelta}-~\ref{ldelta}.

\begin{table*}[htbp]
\caption{\footnotesize The numerical results of the probabilities $\mathcal{P}{q\bar{q}}$
and the matrix elements of the quark spin $\Delta f_{i}^{5q}$ of all the 17 five-quark
configurations of proton in Set I and Set II. Note that we have denoted the five-quark configurations
$uudq\bar{q}$ with light quark-antiquark pairs as $l\bar{l}$, and those with strange
quark-antiquark pairs as $s\bar{s}$.}
\label{pdelta}
\renewcommand
\tabcolsep{0.45cm}
\renewcommand{\arraystretch}{2.0}
\scriptsize
\begin{tabular}{cc|cccc|cccc}
\toprule[1.2pt]

                       &            &    \multicolumn{4}{|c|}{Set I} &  \multicolumn{4}{c}{Set II} \\\hline

Con. $i$               & $q\bar{q}$ & $\mathcal{P}{q\bar{q}} $ & $\Delta u_i^{5q}$ & $\Delta d_i^{5q}$ & $\Delta s_i^{5q}$ & $\mathcal{P}{q\bar{q}}$ & $\Delta u_i^{5q}$ &   $\Delta d_i^{5q}$         & $\Delta s_i^{5q}$ \\
\hline


\multirow{2}{*}{ $1$ } & $l\bar{l}$ & $0.146\pm0.015$ &        $0$     &    $-1/3$      &      $0$       & $0.157\pm0.016$ &   $0$       & $-1/3$                 & $0$            \\

                       & $s\bar{s}$ & $0.010\pm0.001$ &        $0$     &      $0$       &      $-1/3$    & $0.011\pm0.002$ &  $0$        & $0$                    & $-1/3$\\
\hline

\multirow{2}{*}{ $2$ } & $l\bar{l}$ &        $0$      &        $-$     &      $-$       &      $-$       &       $0$       &   $-$       &  $-$                    & $-$ \\

                       & $s\bar{s}$ & $0.004\pm0.001$ &        $0$     &      $0$       &      $-1/3$    & $0.004\pm0.001$  &  $0$       & $0$                    & $-1/3$\\
\hline

\multirow{2}{*}{ $3$ } & $l\bar{l}$ & $0.016\pm0.002$ &       $-2/9$   &     $-1/9$     &        $0$     & $0.018\pm0.002$  &  $-2/9$    & $-1/9$                 & $0$ \\

                       & $s\bar{s}$ &        $0$      &        $-$     &      $-$       &         $-$    &         $0$      &    $-$     &      $-$                    &      $-$  \\
\end{tabular}

\end{table*}


\begin{table*}[htbp]
\caption{\footnotesize The numerical results of the probabilities $\mathcal{P}{q\bar{q}}$
and the matrix elements of the quark spin $\Delta f_{i}^{5q}$ of the neutron, conventions are the same as in
Table~\ref{pdelta}.}
\label{ndelta}
\renewcommand
\tabcolsep{0.45cm}
\renewcommand{\arraystretch}{2.0}
\scriptsize
\begin{tabular}{cc|cccc|cccc}
\toprule[1.2pt]

                       &            &    \multicolumn{4}{|c|}{Set I} &  \multicolumn{4}{c}{Set II} \\\hline

Con. $i$               & $q\bar{q}$ & $\mathcal{P}{q\bar{q}} $ & $\Delta u_i^{5q}$ & $\Delta d_i^{5q}$ & $\Delta s_i^{5q}$ & $\mathcal{P}{q\bar{q}}$ & $\Delta u_i^{5q}$ &   $\Delta d_i^{5q}$         & $\Delta s_i^{5q}$ \\
\hline


\multirow{2}{*}{ $1$ } & $l\bar{l}$ & $0.146\pm0.015$ &        $-1/3$     &    $0$      &      $0$       & $0.157\pm0.016$ &   $-1/3$       & $0$                 & $0$            \\

                       & $s\bar{s}$ & $0.010\pm0.001$ &        $0$     &      $0$       &      $-1/3$    & $0.011\pm0.002$ &  $0$        & $0$                    & $-1/3$\\
\hline

\multirow{2}{*}{ $2$ } & $l\bar{l}$ &        $0$      &        $-$     &      $-$       &      $-$       &       $0$       &   $-$       &  $-$                    & $-$ \\

                       & $s\bar{s}$ & $0.004\pm0.001$ &        $0$     &      $0$       &      $-1/3$    & $0.004\pm0.001$  &  $0$       & $0$                    & $-1/3$\\
\hline

\multirow{2}{*}{ $3$ } & $l\bar{l}$ & $0.016\pm0.002$ &       $-1/9$   &     $-2/9$     &        $0$     & $0.018\pm0.002$  &  $-1/9$    & $-2/9$                 & $0$ \\

                       & $s\bar{s}$ &        $0$      &        $-$     &      $-$       &         $-$    &         $0$      &    $-$     &      $-$                    &      $-$  \\
\end{tabular}

\end{table*}


\begin{table*}[htbp]
\caption{\footnotesize The numerical results of the probabilities $\mathcal{P}{q\bar{q}}$
and the matrix elements of the quark spin $\Delta f_{i}^{5q}$ of $\Sigma^+$, conventions are the same as in
Table~\ref{pdelta}.}
\label{s+delta}
\renewcommand
\tabcolsep{0.45cm}
\renewcommand{\arraystretch}{2.0}
\scriptsize
\begin{tabular}{cc|cccc|cccc}
\toprule[1.2pt]

                       &            &    \multicolumn{4}{|c|}{Set I} &  \multicolumn{4}{c}{Set II} \\\hline

Con. $i$               & $q\bar{q}$ & $\mathcal{P}{q\bar{q}} $ & $\Delta u_i^{5q}$ & $\Delta d_i^{5q}$ & $\Delta s_i^{5q}$ & $\mathcal{P}{q\bar{q}}$ & $\Delta u_i^{5q}$ &   $\Delta d_i^{5q}$         & $\Delta s_i^{5q}$ \\
\hline


\multirow{2}{*}{ $1$ } & $l\bar{l}$ & $0.066\pm0.007$ &        $0$     &    $-1/3$      &      $0$       & $0.071\pm0.008$ &   $0$       & $-1/3$                 & $0$            \\

                       & $s\bar{s}$ & $0.020\pm0.002$ &        $0$     &      $0$       &      $-1/3$    & $0.021\pm0.003$ &  $0$        & $0$                    & $-1/3$\\
\hline

\multirow{2}{*}{ $2$ } & $l\bar{l}$ & $0.010\pm0.001$      &        $0$     &      $-1/3$       &      $0$       &       $0.010\pm0.001$       &   $0$       &  $-1/3$                    & $0$ \\

                       & $s\bar{s}$ & $0$ &        $-$     &      $-$       &      $-$    & $0$  &  $-$       & $-$                    & $-$\\
\hline

\multirow{2}{*}{ $3$ } & $l\bar{l}$ & $0.017\pm0.002$ &       $-1/4$   &     $-1/12$     &        $0$     & $0.018\pm0.002$  &  $-1/4$    & $-1/12$                 & $0$ \\

                       & $s\bar{s}$ & $0.004\pm0.001$      &        $0$     &      $0$       &         $-1/3$    &         $0.004\pm0.001$      &    $0$     &      $0$                    &      $-1/3$  \\
\end{tabular}

\end{table*}


\begin{table*}[htbp]
\caption{\footnotesize The numerical results of the probabilities $\mathcal{P}{q\bar{q}}$
and the matrix elements of the quark spin $\Delta f_{i}^{5q}$ of $\Sigma^0$, conventions are the same as in
Table~\ref{pdelta}.}
\label{s0delta}
\renewcommand
\tabcolsep{0.45cm}
\renewcommand{\arraystretch}{2.0}
\scriptsize
\begin{tabular}{cc|cccc|cccc}
\toprule[1.2pt]

                       &            &    \multicolumn{4}{|c|}{Set I} &  \multicolumn{4}{c}{Set II} \\\hline

Con. $i$               & $q\bar{q}$ & $\mathcal{P}{q\bar{q}} $ & $\Delta u_i^{5q}$ & $\Delta d_i^{5q}$ & $\Delta s_i^{5q}$ & $\mathcal{P}{q\bar{q}}$ & $\Delta u_i^{5q}$ &   $\Delta d_i^{5q}$         & $\Delta s_i^{5q}$ \\
\hline


\multirow{2}{*}{ $1$ } & $l\bar{l}$ & $0.066\pm0.007$ &        $-1/6$     &    $-1/6$      &      $0$       & $0.071\pm0.008$ &   $-1/6$       & $-1/6$                 & $0$            \\

                       & $s\bar{s}$ & $0.020\pm0.002$ &        $0$     &      $0$       &      $-1/3$    & $0.021\pm0.003$ &  $0$        & $0$                    & $-1/3$\\
\hline

\multirow{2}{*}{ $2$ } & $l\bar{l}$ & $0.010\pm0.001$      &        $-1/6$     &      $-1/6$       &      $0$       &       $0.010\pm0.001$       &   $-1/6$       &  $-1/6$                    & $0$ \\

                       & $s\bar{s}$ & $0$ &        $-$     &      $-$       &      $-$    & $0$  &  $-$       & $-$                    & $-$\\
\hline

\multirow{2}{*}{ $3$ } & $l\bar{l}$ & $0.017\pm0.002$ &       $-1/6$   &     $-1/6$     &        $0$     & $0.018\pm0.002$  &  $-1/6$    & $-1/6$                 & $0$ \\

                       & $s\bar{s}$ & $0.004\pm0.001$      &        $0$     &      $0$       &         $-1/3$    &         $0.004\pm0.001$      &    $0$     &      $0$                    &      $-1/3$  \\
\end{tabular}

\end{table*}


\begin{table*}[htbp]
\caption{\footnotesize The numerical results of the probabilities $\mathcal{P}{q\bar{q}}$
and the matrix elements of the quark spin $\Delta f_{i}^{5q}$ of $\Sigma^-$, conventions are the same as in
Table~\ref{pdelta}.}
\label{s-delta}
\renewcommand
\tabcolsep{0.45cm}
\renewcommand{\arraystretch}{2.0}
\scriptsize
\begin{tabular}{cc|cccc|cccc}
\toprule[1.2pt]

                       &            &    \multicolumn{4}{|c|}{Set I} &  \multicolumn{4}{c}{Set II} \\\hline

Con. $i$               & $q\bar{q}$ & $\mathcal{P}{q\bar{q}} $ & $\Delta u_i^{5q}$ & $\Delta d_i^{5q}$ & $\Delta s_i^{5q}$ & $\mathcal{P}{q\bar{q}}$ & $\Delta u_i^{5q}$ &   $\Delta d_i^{5q}$         & $\Delta s_i^{5q}$ \\
\hline


\multirow{2}{*}{ $1$ } & $l\bar{l}$ & $0.066\pm0.007$ &        $-1/3$     &    $0$      &      $0$       & $0.071\pm0.008$ &   $-1/3$       & $0$                 & $0$            \\

                       & $s\bar{s}$ & $0.020\pm0.002$ &        $0$     &      $0$       &      $-1/3$    & $0.021\pm0.003$ &  $0$        & $0$                    & $-1/3$\\
\hline

\multirow{2}{*}{ $2$ } & $l\bar{l}$ & $0.010\pm0.001$      &        $-1/3$     &      $0$       &      $0$       &       $0.010\pm0.001$       &   $-1/3$       &  $0$                    & $0$ \\

                       & $s\bar{s}$ & $0$ &        $-$     &      $-$       &      $-$    & $0$  &  $-$       & $-$                    & $-$\\
\hline

\multirow{2}{*}{ $3$ } & $l\bar{l}$ & $0.017\pm0.002$ &       $-1/12$   &     $-1/4$     &        $0$     & $0.018\pm0.002$  &  $-1/12$    & $-1/4$                 & $0$ \\

                       & $s\bar{s}$ & $0.004\pm0.001$      &        $0$     &      $0$       &         $-1/3$    &         $0.004\pm0.001$      &    $0$     &      $0$                    &      $-1/3$  \\
\end{tabular}

\end{table*}


\begin{table*}[htbp]
\caption{\footnotesize The numerical results of the probabilities $\mathcal{P}{q\bar{q}}$
and the matrix elements of the quark spin $\Delta f_{i}^{5q}$ of $\Xi^0$, conventions are the same as in
Table~\ref{pdelta}.}
\label{x0delta}
\renewcommand
\tabcolsep{0.45cm}
\renewcommand{\arraystretch}{2.0}
\scriptsize
\begin{tabular}{cc|cccc|cccc}
\toprule[1.2pt]
                       &            &    \multicolumn{4}{|c|}{Set I} &  \multicolumn{4}{c}{Set II} \\\hline

Con. $i$               & $q\bar{q}$ & $\mathcal{P}{q\bar{q}} $ & $\Delta u_i^{5q}$ & $\Delta d_i^{5q}$ & $\Delta s_i^{5q}$ & $\mathcal{P}{q\bar{q}}$ & $\Delta u_i^{5q}$ &   $\Delta d_i^{5q}$         & $\Delta s_i^{5q}$ \\
\hline


\multirow{2}{*}{ $1$ } & $l\bar{l}$ & $0.083\pm0.009$ &        $-2/9$     &    $-1/9$      &      $0$       & $0.094\pm0.011$ &   $-2/9$       & $-1/9$                 & $0$            \\

                       & $s\bar{s}$ & $0$ &        $-$     &      $-$       &      $-$    & $0$ &  $-$        & $-$                    & $-$\\
\hline

\multirow{2}{*}{ $2$ } & $l\bar{l}$ & $0.010\pm0.001$      &        $0$     &      $-1/3$       &      $0$       &       $0.011\pm0.001$       &   $0$       &  $-1/3$                    & $0$ \\

                       & $s\bar{s}$ & $0$ &        $-$     &      $-$       &      $-$    & $0$  &  $-$       & $-$                    & $-$\\
\hline

\multirow{2}{*}{ $3$ } & $l\bar{l}$ & $0.011\pm0.002$ &       $-2/9$   &     $-1/9$     &        $0$     & $0.012\pm0.001$  &  $-2/9$    & $-1/9$                 & $0$ \\

                       & $s\bar{s}$ & $0.009\pm0.001$      &        $0$     &      $0$       &         $-1/3$    &         $0.010\pm0.002$      &    $0$     &      $0$                    &      $-1/3$  \\
\end{tabular}

\end{table*}

\begin{table*}[htbp]
\caption{\footnotesize The numerical results of the probabilities $\mathcal{P}{q\bar{q}}$
and the matrix elements of the quark spin $\Delta f_{i}^{5q}$ of $\Xi^-$, conventions are the same as in
Table~\ref{pdelta}.}
\label{x-delta}
\renewcommand
\tabcolsep{0.45cm}
\renewcommand{\arraystretch}{2.0}
\scriptsize
\begin{tabular}{cc|cccc|cccc}
\toprule[1.2pt]

                       &            &    \multicolumn{4}{|c|}{Set I} &  \multicolumn{4}{c}{Set II} \\\hline

Con. $i$               & $q\bar{q}$ & $\mathcal{P}{q\bar{q}} $ & $\Delta u_i^{5q}$ & $\Delta d_i^{5q}$ & $\Delta s_i^{5q}$ & $\mathcal{P}{q\bar{q}}$ & $\Delta u_i^{5q}$ &   $\Delta d_i^{5q}$         & $\Delta s_i^{5q}$ \\
\hline


\multirow{2}{*}{ $1$ } & $l\bar{l}$ & $0.083\pm0.009$ &        $-1/9$     &    $-2/9$      &      $0$       & $0.094\pm0.011$ &   $-1/9$       & $-2/9$                 & $0$            \\

                       & $s\bar{s}$ & $0$ &        $-$     &      $-$       &      $-$    & $0$ &  $-$        & $-$                    & $-$\\
\hline

\multirow{2}{*}{ $2$ } & $l\bar{l}$ & $0.010\pm0.001$      &        $-1/3$     &      $0$       &      $0$       &       $0.011\pm0.001$       &   $-1/3$       &  $0$                    & $0$ \\

                       & $s\bar{s}$ & $0$ &        $-$     &      $-$       &      $-$    & $0$  &  $-$       & $-$                    & $-$\\
\hline

\multirow{2}{*}{ $3$ } & $l\bar{l}$ & $0.011\pm0.002$ &       $-1/9$   &     $-2/9$     &        $0$     & $0.012\pm0.001$  &  $-1/9$    & $-2/9$                 & $0$ \\

                       & $s\bar{s}$ & $0.009\pm0.001$      &        $0$     &      $0$       &         $-1/3$    &         $0.010\pm0.002$      &    $0$     &      $0$                    &      $-1/3$  \\
\end{tabular}

\end{table*}

\begin{table*}[htbp]
\caption{\footnotesize The numerical results of the probabilities $\mathcal{P}{q\bar{q}}$
and the matrix elements of the quark spin $\Delta f_{i}^{5q}$ of $\Lambda$, conventions are the same as in
Table~\ref{pdelta}.}
\label{ldelta}
\renewcommand
\tabcolsep{0.45cm}
\renewcommand{\arraystretch}{2.0}
\scriptsize
\begin{tabular}{cc|cccc|cccc}
\toprule[1.2pt]

                       &            &    \multicolumn{4}{|c|}{Set I} &  \multicolumn{4}{c}{Set II} \\\hline

Con. $i$               & $q\bar{q}$ & $\mathcal{P}{q\bar{q}} $ & $\Delta u_i^{5q}$ & $\Delta d_i^{5q}$ & $\Delta s_i^{5q}$ & $\mathcal{P}{q\bar{q}}$ & $\Delta u_i^{5q}$ &   $\Delta d_i^{5q}$         & $\Delta s_i^{5q}$ \\
\hline


\multirow{2}{*}{ $1$ } & $l\bar{l}$ & $0.115\pm0.013$ &        $-1/6$     &    $-1/6$      &      $0$       & $0.128\pm0.014$ &   $-1/6$       & $-1/6$                 & $0$            \\

                       & $s\bar{s}$ & $0$ &        $-$     &      $-$       &      $-$    & $0$ &  $-$        & $-$                    & $-$\\
\hline

\multirow{2}{*}{ $2$ } & $l\bar{l}$ & $0.003\pm0.001$      &        $-1/6$     &      $-1/6$       &      $0$       &       $0.003\pm0.000$       &   $-1/6$       &  $-1/6$                    & $0$ \\

                       & $s\bar{s}$ & $0.003\pm0.001$ &        $0$     &      $0$       &      $-1/3$    & $0.003\pm0.000$  &  $0$       & $0$                    & $-1/3$\\
\hline

\multirow{2}{*}{ $3$ } & $l\bar{l}$ & $0$ &       $-$   &     $-$     &        $-$     & $0$  &  $-$    & $-$                 & $-$ \\

                       & $s\bar{s}$ & $0$      &        $-$     &      $-$       &         $-$    &         $0$      &    $-$     &      $-$                    &      $-$  \\
\end{tabular}

\end{table*}

\end{appendix}

\end{document}